\documentstyle[12pt]{article}

\makeatletter
\def\section{\@startsection{section}{1}{\z@}{-3.5ex plus -1ex minus -.2ex}
{2.3ex plus .2ex}{\large\bf}}
\makeatother
 
\makeatletter
\@addtoreset{equation}{section}

\makeatother

\def\lb{\lbrack}
\def\rb{\rbrack}

 \def\Slash#1{
  \begin{picture}(5,6)(0,0)
  \put(-.7,-1.2){\line(5,6)6}
  \end{picture}
  \kern-.8em#1}
 \def\slash#1{
  \begin{picture}(5,6)(0,0)
  \put(-1.5,-1.7){\line(5,6)5}
  \end{picture}
  \kern-.8em#1}

\def\sd{\Slash \partial}

\def\g5{\gamma_5}
\def\hg5{\hat{\gamma}_5}
\def\A{{\cal A}}

\def\C{{\cal C}}

\def\G{{\cal G}}
\def\U{{\cal U}}

\def\S{{\cal S}}
\def\N{{\cal N}}
\def\P{{\cal P}}

\def\hC{\widehat{\cal C}}

\def\tA{A_f}
\def\tPhi{\widetilde{\Phi}}
\def\tlambda{\widetilde{\lambda}}

\def\Qlatmr1{Q_{lat}^{(m=r=1)}}

\def\be{\begin{eqnarray}}
\def\ee{\end{eqnarray}}

\def\SU{\mbox{SU}}

\parskip=\medskipamount

\def\qed{\vrule height 1.5ex width 1.2ex depth -.1ex}

\begin{document}

\input epsf

\vspace{8mm}

\begin{center}
 
{\Large \bf Gauge fixing, families index theory, and topological features 
of the space of lattice gauge fields}
\\

\vspace{0.3ex}

\vspace{12mm}

{\large David H. Adams}

\vspace{4mm}

Physics dept., National Taiwan University, Taipei, Taiwan 106, R.O.C.\\

and \\

Physics dept., Duke University, Durham, NC 27708, U.S.A.\footnote{Current 
address}\\

\vspace{1ex}

email: adams@phy.duke.edu

\end{center}

\begin{abstract}

The families index theory for the overlap lattice Dirac operator is 
applied to derive topological features of the space of SU($N$) lattice gauge 
fields on the 4-torus: The topological sectors, specified by the fermionic
topological charge, are shown to contain noncontractible even-dimensional
spheres when $N\ge3$, and noncontractible circles in the $N\!=\!2$ case.
We describe how certain obstructions to the existence of gauge
fixings without the Gribov problem in the continuum setting correspond
on the lattice to obstructions to the contractibility of these spheres and 
circles. We also point out a canonical connection on the space of lattice
gauge fields with monopole-like singularities associated with the spheres.

\end{abstract}

\section{Introduction}

Lattice gauge fields on the 4-torus $T^4$ have a fermionic topological 
charge which arises in the overlap formalism \cite{ov} and can be expressed
as the index of the overlap Dirac operator \cite{Neu1}. It determines a 
decomposition of the space $\U$ of lattice gauge fields 
(with a given unitary gauge
group) into topological sectors after excluding a measure-zero subspace
on which the fermionic topological charge is
ill-defined. In this paper we consider the following natural question: 
what do these topological sectors look like? Their topology has been 
worked out in the U(1) case (with a certain admissibility condition
imposed on the gauge fields) by L\"uscher in \cite{L(abelian)}. 
This was a key part of the existence proof
in \cite{L(abelian)} for gauge invariant abelian chiral gauge theory on the
lattice when anomalies cancel.\footnote{This result, which has also 
been established by different 
means in the noncompact U(1) case \cite{Neu(noncompactU(1))},
generalises to U(1) chiral gauge theory on arbitrary even-dimensional torus 
\cite{Suzuki(general)}, and to the electro-weak U(1)$\times$SU(2) case 
\cite{Kiku}. In the general 
nonabelian case the existence of chiral gauge theory with exact gauge 
invariance on the lattice has been shown at the perturbative level in 
\cite{Suzuki,L(pert)}. However, despite the progress in \cite{L(nonabelian)},
there is at present no nonperturbative existence proof (except for the special
case where the fermion representation is real \cite{Suzuki(real)}).}
In this paper
we derive first results on the topology of the sectors of $\U$ in the 
nonabelian SU($N$) case. Besides copies of SU($N$), which are
present because the lattice gauge fields take values in the gauge group, 
we show that the sectors of $\U$ contain certain noncontractible 
even-dimensional spheres in the $N\ge3$ case, and noncontractible circles
in the $N\!=\!2$ case. These are a direct consequence of excluding the 
lattice gauge fields for which the fermionic topological charge is 
ill-defined.

The presence of the noncontractible $2n$-spheres in $\U$ is derived  
from results on the orbit space of $\U$ obtained 
via the families index theory for the overlap Dirac operator in 
\cite{DA(I),DA(II)}. The classical continuum limit result Theorem 2 of 
\cite{DA(I)} reveals the presence of certain 
noncontractible $2n$-spheres in the orbit space in the $N\ge3$ case.
Similarly, the classical continuum limit result 
of Ref. \cite{Bar} on the lattice
version of Witten's global anomaly implies the presence of noncontractible
circles in the orbit space of $\U$ in the SU(2) case. 
We show that these 
give rise to noncontractible $2n$-spheres/circles in $\U$ itself by 
exploiting the known fact that, in contrast to the continuum situation, 
gauge fixings without the Gribov problem exist on the lattice 
(e.g., the maximal tree gauges \cite{Creutz,DeTar}). 
This is done in Section 2.
The argument is quite implicit though, and our aim in the rest of
the paper is to give a more explicit and illuminating construction 
of noncontractible $2n$-spheres/circles in $\U\,$, and explain how 
the obstructions to their contractibility are the 
lattice counterparts of certain obstructions to the existence of gauge fixings 
without the Gribov problem in the continuum setting. We begin in Section 3
by discussing how the gauge fixing issue and topology of the gauge field 
sectors get related, both in the continuum and on the lattice, when 
considering the gauge invariance issue for the chiral fermion determinant.
In Section 4, we give a fermionic description of certain obstructions to
the existence of Gribov problem-free gauge fixings in the continuum SU($N$)
gauge theory, based on families index theory for the Dirac operator, and
Witten's global anomaly, in the $N\ge3$ and $N\!=\!2$ cases, respectively. 
The lattice version of these considerations in Section 5 leads instead to
obstructions to the contractibility of $2n$-spheres/circles in $\U$.
We also discuss a connection on the space of lattice gauge fields, with 
monopole-like singularities associated with the $2n$-spheres, which arises
naturally in this context.
The results of the paper are summarised in Section 6. 
A property of $\G_0$ gauge transformations used in the text is derived
in an appendix.

\section{Topology of the space of lattice gauge fields}

We assume familiarity with the lattice formulation of SU($N$) gauge theory
on $T^4$ as summarised in \cite{DA(I)} and refer to that paper for the 
definitions and notations used in the following. To begin with, the space
of lattice gauge fields is
\be
\U_{\rm initial}=\SU(N)\times\SU(N)\times\cdots\times\SU(N)
\label{1}
\ee
(one copy for each lattice link). This space is connected, but decomposes
into disconnected sectors labeled by the fermionic topological charge
after excluding the codimension 1 submanifold of gauge fields for which the
charge is ill-defined. The fermionic topological charge is given by the 
index of the overlap Dirac operator 
\cite{Neu1}\footnote{It reduces to the continuum topological charge in the 
classical continuum limit \cite{DA(JMP)}.},
and the excluded fields $U$ are those for which this operator is ill-defined, 
i.e. the ones for which the Hermitian Wilson-Dirac operator $H^U$
(with suitable negative mass term) has
zero-modes. We denote the resulting space of lattice gauge fields by $\U$.
A sufficient (although not necessary) condition for a
lattice gauge field $U$ to lie in $\U$ is that its plaquette
variables satisfy the ``admissibility condition'',
\be
||1-U(p)||\;<\;\epsilon
\label{2}
\ee
where $U(p)$ is the product of the $U_{\mu}(x)$'s around the lattice plaquette
$p$. For sufficiently small $\epsilon$, this condition guarantees the absence
of zero-modes for $H^U$ \cite{L(local),Neu(bound)}. When $U$ is the lattice
transcript of a continuum field $A$, and $p$ is the plaquette specified by 
a lattice site $x$ and directions $\mu$ and $\nu$, then
$1-U(p)=a^2F_{\mu\nu}(x)+O(a^3)$. Hence the lattice transcript of a 
smooth continuum field (or family of continuum fields) is guaranteed to lie 
in $\U$ when the lattice is sufficiently fine (see \cite{DA(JMP)} for further
discussion of this point). 

Since $\U_{\rm initial}$ is a product of copies of SU($N$),
its topology is known. The topology of $\U$ is more 
complicated and has not yet been determined in the present SU($N$) case.
(It has so far only been worked out in the U(1) case with the 
admissibility condition (\ref{2}) imposed \cite{L(abelian)}; we discuss 
this case further below.) In the following we apply results from the 
families index theory for the overlap Dirac operator \cite{DA(I)} to derive
first results on the topology of $\U$ in the SU($N$) case:

\vspace{1ex}

\noindent {\bf Theorem}. When the lattice is sufficiently fine, the
topological sectors of $\U$   
contain noncontractible $2n$-dimensional spheres for $1\le{}n\le{}N\!-\!2$
in the $N\ge3$ case, and noncontractible circles in the SU(2) case.

\vspace{1ex}

\noindent {\bf Remark}. Note that, by (\ref{1}), 
$\pi_{2n}(\U_{\rm initial})=0$ for $n\le{}N\!-\!1$ and 
$\pi_1(\U_{\rm initial})=0\,$, since the same is true for $\pi_{2n}$ and 
$\pi_1$ of $\SU(N)$. In contrast, by the theorem, $\pi_{2n}(\U)\ne0$
for $1\le{}n\le{}N\!-\!2$ in the $N\ge3$ case, and $\pi_1(\U)\ne0$
in the SU(2) case. 
Hence the noncontractible $2n$-spheres and circles mentioned in the theorem
are all contractible in $\U_{\rm initial}\,$; their 
noncontractibility in $\U$ reflects the topological consequences of excluding
from $\U_{\rm initial}$ the fields $U$ for which the fermionic topological
charge is ill-defined.

\vspace{1ex}

We will give two proofs of the Theorem in this paper. Both involve studying
$\U$ as a $\G_0$-bundle over the orbit space $\U/\G_0\,$, where $\G_0$
is the subgroup of gauge transformations 
$\phi:\{\mbox{lattice sites}\}\to\SU(N)$ satisfying the condition 
$\phi(x_0)=1$ for some arbitrarily chosen basepoint $x_0$ in $T^4$. 
The reason for
this condition is to exclude the (nontrivial) constant gauge transformations;
consequently, $\G_0$ acts freely on $\U$ (unlike the full 
group $\G$).\footnote{For example, the trivial gauge field $U\!=\!1$ is
invariant under all constant gauge transformations, and constant $\phi$
with values in the center of $\SU(N)$ act trivially on all gauge fields.}
This is crucial for our arguments in the following. The first, most direct  
proof of the theorem is given in the remainder of this Section.
It makes use of the fact that $\G_0$ gauge fixings without the Gribov 
problem exist on the lattice. Examples of these are the maximal tree gauges 
\cite{Creutz,DeTar} which we review further below.
Gauge fixings which do not have the Gribov problem \cite{Gribov} 
are referred to as `good' in the following.
A good $\G_0$ gauge fixing picks out a submanifold $\U_f$ of $\U$ which 
intersects each $\G_0$ orbit precisely once. Since $\G_0$ acts freely on
$\U$, this determines a decomposition
\be
\U\;\simeq\;\U_f\times\G_0
\label{3}
\ee
The one-to-one correspondence between elements of $\U_f$ and $\U/\G_0$
then gives
\be
\U\;\simeq\;\U/\G_0\times\G_0
\label{4}
\ee
i.e. a trivialisation of $\U$ as a $\G_0$-bundle over $\U/\G_0$.
Hence the topology of $\U$ is determined by that of $\U/\G_0$ and $\G_0$.
The $N\ge3$ part of the Theorem now follows from results obtained via
the families index theory for the overlap Dirac operator in \cite{DA(I)}:
It was shown there that the topological sectors of $\U/\G_0$ contain
noncontractible $2n$-spheres for $1\le{}n\le{}N\!-\!2\,$; these arise as the 
lattice transcripts of certain $2n$-spheres in the continuum orbit space
$\A/\G_0$ over which the topological charge (integrated Chern character) 
of the index bundle of the continuum Dirac operator is nonvanishing.
By Theorem 2 of \cite{DA(I)}, the topological charges of the lattice index 
bundle over the lattice transcripts of these $2n$-spheres coincide with the 
continuum topological charges when the lattice is sufficiently fine.
Since the charges are constant under smooth deformations of the spheres, 
their nonvanishing implies noncontractibility of these $2n$-spheres in 
$\U/\G_0$. This together with the decomposition (\ref{4}) implies the presence
of the noncontractible $2n$-spheres in $\U$ itself, as claimed in the Theorem.

The presence of noncontractible circles in $\U$ in the SU(2) case can be 
derived in a similar way, using the results of Ref. \cite{Bar} on the lattice
version of Witten's global anomaly.\footnote{The presence of the global
anomaly on the lattice had earlier been verified numerically in
\cite{Neu(SU(2)),Bar1}.} 
The lattice version of the global anomaly
can be expressed as $\N$ (mod 2), where $\N$ is the net number of crossings 
of the origin (counted with sign) by the eigenvalues of the overlap Dirac 
operator as the background gauge field is smoothly varied along a path
$U_t$ from an initial field $U_0$ to a final field $U_1=\phi\cdot{}U_0$.
The results of \cite{Bar} show that, when $U_t$ and $\phi$ are the lattice 
transcripts of a continuum path $A_t$ and a topologically nontrivial map
$\phi:T^4\to\SU(2)\,$, then the lattice anomaly reproduces the continuum one,
i.e. $\N$ (mod 2)$=1$ when the lattice is sufficiently fine (for a single
fermion species in the fundamental representation of SU(2)).\footnote{
The requirement that $\phi(x_0)=1$ is inconsequential here since we can 
replace $\phi\to\phi(x_0)^{-1}\phi$ to enforce this condition without 
affecting the topological properties of $\phi$.}
The path $U_t$ is a circle in $\U/\G_0\,$, and clearly $\N$ can only change
by even integers under deformations of this circle. Hence the nonvanishing of
$\N$ (mod 2) implies noncontractibility of the circle in $\U/\G_0$.
This together with (\ref{4}) implies the presence of noncontractible circles 
in $\U$ itself. This completes our first proof of the Theorem.

The preceding argument gives more information on $\pi_{2n}(\U)$ besides the 
fact that it is nonvanishing for $1\le{}n\le{}N\!-\!2$: The topological 
charges of the $2n$-spheres in the continuum orbit space $\A/\G_0$ discussed
above can have arbitrary integer values; hence for any given integer $p\,$,
$\U/\G_0$ will contain a $2n$-sphere with topological charge $p$ when the 
lattice is sufficiently fine. Since $2n$-spheres in $\U/\G_0$ with 
different topological charges represent different elements in 
$\pi_{2n}(\U/\G_0)$, it follows that the number of different elements in
$\pi_{2n}(\U/\G_0)$ can be made arbitrarily large by taking the lattice to be
sufficiently fine, and becomes infinite in the continuum limit.
By (\ref{4}), the same is true for $\pi_{2n}(\U)$. On the other hand, the 
argument in the SU(2) case does not indicate more than one nontrivial element
in $\pi_1(\U/\G_0)$ or $\pi_1(\U)$ since there is only one equivalence class
of topologically nontrivial maps $\phi:T^4\to\SU(2)$. 

We remark that imposing the admissibility condition (\ref{2}) does not change 
the situation regarding the topological features of $\U$ derived above.
The $2n$-spheres/circles in $\U/\G_0\,$, from which the $2n$-spheres/circles
in $\U$ arise, are the lattice transcripts of certain $2n$-spheres/circles
in $\A/\G_0$ which come from $2n$-dimensional balls/line segments 
in $\A$ (see \cite{DA(I)}), and the lattice transcripts of such families
of continuum gauge fields are guaranteed to satisfy the admissibility
condition for any $\epsilon>0$ when the lattice is sufficiently fine,
cf. the discussion following (\ref{2}) above. Then the arguments above go 
through unchanged.

\begin{figure}
$$
\epsfysize=4cm \epsfbox{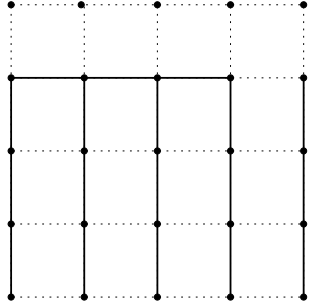}
$$
\caption{A maximal tree (the solid links) in a $4\times4$ lattice.}
\label{gftl}
\end{figure}

The decomposition (\ref{3}), which was crucial for the above proof of the
Theorem, relies on the fact that good
$\G_0$ gauge fixings exist on the lattice. Such gauge fixings can be obtained
from maximal trees \cite{Creutz,DeTar} as we now discuss. A {\em tree} in
the lattice is a collection of lattice links from which no closed loops
can be formed. The tree is called {\em maximal} if it is not possible to
add another lattice link without getting a closed loop. An example of a 
maximal tree in a 2-dimensional lattice is given in Fig.\ \ref{gftl}.
It was pointed out by Creutz \cite{Creutz} (see also \cite{DeTar})
that, given a maximal tree, any lattice gauge field can be gauge transformed
into a field whose link variables along the tree are all equal to 1 in a way
that is unique up to constant gauge transformations. Furthermore, any two
gauge fields in the same gauge orbit get transformed into fields which 
coincide up to a constant gauge transformation. Thus a maximal tree determines
a lattice gauge fixing which, modulo constant gauge transformations, is free
of Gribov ambiguities.\footnote{As discussed in \cite{Mandula},  
the residual symmetry  under constant gauge transformations is easily handled. 
(General methods for dealing with theories with a residual global symmetry 
have also been discussed in \cite{lor}.)  However, finding
physically acceptable lattice gauge fixings which can be implemented in
Monte Carlo simulations without Gribov copies arising is a nontrivial
problem which has yet to be completely resolved. For reviews of this issue
see, e.g., \cite{gfreview,vBaal}.}  
We now point out that maximal trees determine good $\G_0$
gauge fixings. Indeed, for each lattice gauge field there is a unique 
gauge transformation satisfying $\phi(x_0)=1$ which transforms it to a field
whose link variables along the tree are all equal to 1. The key observation
now is that any two gauge fields in the same $\G_0$ orbit get sent to  
the same gauge-fixed field under this procedure. This is seen as follows.
For each link not contained in the maximal tree, the link variable of the 
gauge-fixed field is the product of the original link variables around a 
closed loop, starting and ending at $x_0$, which is formed by adding the link 
to the tree. (We leave the straightforward verification of this fact to the 
interested reader.) Consequently, under a $\G_0$ gauge transformation of the 
original field, the nontrivial link variables of the gauge-fixed field
just get conjugated by $\phi(x_0)=1$, i.e. they are unchanged. 
Thus the procedure picks out precisely one lattice gauge field in each $\G_0$
orbit, i.e. we have a good $\G_0$ gauge fixing.
 
We note in passing the topological structure of $\U_{\rm initial}/\G_0\,$,
which can be determined from a maximal tree gauge fixing as follows.
The ``gauge-fixed'' submanifold $\U_{f({\rm initial})}$ of $\U_{\rm initial}$
picked out by a maximal tree $\G_0$ gauge fixing consists of the lattice
gauge fields whose link variables are 1 on the links of the tree. 
I.e. $\U_{f({\rm initial})}$ is a product of copies of SU($N$) with one copy
for each lattice link not contained in the tree. A general property of maximal
trees (whose straightforward verification we again leave to the reader)
is that the number of links making up the tree is $s-1$, where $s$
is the number of lattice sites. Thus $\U_{f({\rm initial})}$ is the product
of $4s-(s-1)=3s+1$ copies of SU($N$). 
Since $\U_{f({\rm initial})}\simeq\U_{\rm initial}/\G_0\,$, this reveals the 
topology of $\U_{\rm initial}/\G_0$. In particular, it follows that the 
noncontractible $2n$-spheres/circles in $\U/\G_0$ discussed above are
all contractible in $\U_{\rm initial}/\G_0$. As a small consistency check, 
we also note that the decomposition 
\be
\U_{\rm initial}\;\simeq\;\U_{\rm initial}/\G_0\times\G_0
\label{5}
\ee
reproduces (\ref{1}), since $\G_0$ is the product of $s-1$ copies of 
SU($N$) (one for each lattice site except $x_0$).

In the U(1) case, the topology of $\U$ when the admissibility condition
(\ref{2}) is imposed was completely worked out by L\"uscher in 
\cite{L(abelian)}. In this case, the topological sectors are labeled by 
topological fluxes $m=\{m_{\mu\nu}\}$ rather than a topological charge.
Using a good $\G_0$ gauge fixing (different from the maximal tree gauges;
it is a lattice version of the Landau gauge supplemented with an additional
condition to exclude Gribov copies), L\"uscher showed that the topological
structure of each topological sector is 
\be
\U_{\lb{}m\rb}\;\simeq\;{\rm U}(1)^4\times\U_{\lb{}m\rb}^{\rm cr}
\,\times\,\G_0
\label{7}
\ee
where $\U_{\lb{}m\rb}^{\rm cr}$ is a contractible submanifold in 
$\U_{\lb{}m\rb}$. The ${\rm U}(1)^4$ factor in the gauge-fixed submanifold
${\rm U}(1)^4\times\U_{\lb{}m\rb}^{\rm cr}$ is a remnant of the product
of copies of U(1) making up $\U_{f({\rm initial})}$.
Thus, in the U(1) case, decomposing $\U_{\rm initial}$ into topological
sectors by imposing the admissibility condition (\ref{2})
has the effect of removing some of the initial topological structure
(by ``breaking open'' some of the copies of U(1) to get the contractible
subspace $\U_{\lb{}m\rb}^{\rm cr}$), but does not produce any new nontrivial
topological structure. In contrast, the result of this section shows that,
in the SU($N$) cases, new nontrivial structures are in fact produced
(i.e. the noncontractible $2n$-spheres/circles). 

The existence proof for the noncontractible $2n$-spheres/circles in $\U$
given in this section was rather implicit. In Section 5 we give an 
alternative, more illuminating proof, involving a different, more explicit
construction of $2n$-spheres/circles. In that approach, the obstructions 
to contractibility are seen to be the lattice counterparts of certain 
obstructions to the existence of good $\G_0$ gauge fixings in the smooth
continuum setting (discussed in Section 4). But first, in the next section,
we describe a simpler version of this correspondence that arises naturally 
when considering the gauge invariance issue for the chiral fermion 
determinant.

\section{Relating gauge fixing and topology via the chiral fermion 
determinant}

An important issue in lattice chiral gauge theory is 
whether a smooth, gauge invariant phase choice exists for the (left- or
right-handed) overlap chiral fermion determinant \cite{ov}.
The latter can be 
expressed as $\det(D_{\pm}^U)\,$, where $D$ is the overlap Dirac operator
\cite{Narayanan-Kiku(overlap)}.   
Existence of good $\G_0$ gauge fixings has implications for this: 
If a smooth phase choice for $\det(D_{\pm}^U)$ exists on the gauge-fixed 
submanifold $\U_f$ picked out by the gauge fixing, then a smooth 
{\em gauge invariant} (under $\G_0$)
phase choice on the whole of $\U$ is obtained by simply setting
$\det(D_{\pm}^{\phi\cdot{}U_f})=\det(D_{\pm}^{U_f})$, where $\phi\cdot{}U_f$
denotes the action of a gauge transformation $\phi\in\G_0$ on a gauge field
$U_f\in\U_f$.\footnote{The ``smoothness'' parts of these statements  
break down if $\G_0$ is replaced by $\G$, since the latter does not act
freely on the space of gauge fields.} 

On the other hand it is known, both in the smooth continuum setting
\cite{AG} and in the lattice setting \cite{Neu(geom),DA(NPB),DA(PRL)}, that
there are topological obstructions to gauge invariance of the chiral fermion
determinant in the U(1) and SU($N$) cases (with fermion in the fundamental
representation, or, more generally, when anomalies don't 
cancel).\footnote{The fact that we are restricting the gauge transformations 
to $\G_0$ does not change this situation, cf. the Appendix.} 
It follows that, in these cases, either no good $\G_0$ gauge fixing exists,
or, if one does exist, then no smooth phase choice for the
chiral fermion determinant exists on the submanifold $\U_f$ picked out
by the gauge fixing.

Recall that the chiral fermion determinant is really a section in a U(1)
determinant line bundle over the space of gauge fields (cf. the final section
of \cite{AG} in the continuum, and \cite{ov,Neu(geom),L(abelian),DA(NPB)}
in the lattice setting). A smooth phase choice for the determinant on a 
submanifold of the space 
of gauge fields is equivalent to a trivialisation of the determinant line
bundle over the submanifold. Therefore, the question of whether such a phase 
choice exists is intimately related to the topology of the space of gauge 
fields. In the continuum setting, the topological sectors of the space $\A$
of smooth continuum gauge fields have no nontrivial topological structure --
they are just infinite-dimensional affine vectorspaces. Consequently, by
a standard mathematical fact, the determinant line bundle is trivialisable
over $\A$ (and any submanifold of $\A$). Combining this with the preceding
observations, we conclude that, in the continuum setting, obstructions to
$\G_0$ gauge invariance of the chiral fermion determinant are also 
obstructions to the existence of good $\G_0$ gauge fixings.
In particular, such gauge fixings cannot exist in the U(1) and SU($N$)
continuum gauge theories.\footnote{There are more direct ways to see
these obstructions to $\G_0$ gauge fixings, which do not involve the chiral 
fermion determinant; these are described in the SU($N$) case in Section 3.}

On the other hand, in the lattice setting we already know that good $\G_0$
gauge fixings exist. Then the preceding observations lead to the conclusion
that, when obstructions to gauge invariance are present (e.g., in the U(1)
and SU($N$) cases), the submanifold $\U_f$ picked out by the gauge fixing
is {\em noncontractible} in $\U$. For if $\U_f$ was contractible, then, by a
standard mathematical fact, the determinant line bundle would be trivialisable
over $\U_f\,$, i.e. a smooth phase choice for the chiral fermion determinant
would exist on $\U_f\,$, and we could then obtain a $\G_0$-gauge invariant 
phase choice on the whole of $\U$ in the way described earlier.

The preceding gives a first demonstration of how obstructions to the existence
of good $\G_0$ gauge fixings in the continuum correspond on the lattice
to obstructions to contractibility of certain submanifolds in $\U$.
Actually, the arguments above only establish this correspondence for the
trivial topological sector, since in the other sectors the chiral fermion 
determinant vanishes.
The obstructions to gauge invariance of the chiral fermion determinant have
a natural description in the context of families index theory: they are the
topological charges of the Dirac index bundle over 2-spheres in the gauge 
field orbit space (cf. \cite{DA(I)} and the last section of \cite{AG}
for the lattice and continuum settings, respectively). In the following
sections we use the families index theory to to show more precisely
the correspondence between gauge fixing obstructions in the continuum and 
topological structure in the sectors of the space $\U$ of lattice gauge fields;
the derived results hold in general and not just for the topologically trivial
sector.

Finally, we remark that the noncontractibility of $\U_f$ in the U(1) case,
derived above, is in agreement with L\"uscher's explicit determination of 
$\U_f$ for a particular good $\G_0$ gauge fixing in \cite{L(abelian)}:
From (\ref{7}) we see that each sector of that $\U_f$ is noncontractible
since it contains the noncontractible factor ${\rm U}(1)^4\simeq{}T^4$.

\section{Continuum considerations}

In this section we describe obstructions to good $\G_0$ gauge fixings
in the space $\A$ of smooth continuum SU($N$) gauge fields on $T^4$.
For simplicity we restrict our considerations to the topologically
trivial sector. Then $\G$ consists of the smooth maps 
$\phi:T^4\to\SU(N)$ and $\G_0$ is the subgroup with $\phi(x_0)=1$.
The most direct way to see the obstructions is via the approach 
of I. Singer in 
\cite{Singer} (see also \cite{Syracuse}, where Singer's argument for SU(2)
gauge fields on $S^3$ is generalised to gauge groups including general
SU($N$) and spacetimes including $T^4$):\footnote{Note that our setup is
different from that of \cite{Singer,Syracuse}: There the considerations are
restricted to the {\em irreducible} gauge fields, which are acted freely upon 
by $\G/{\bf Z}_N$ (where ${\bf Z}_N$ is the center of SU($N$)). The 
obstructions are then given by $\pi_n(\G/{\bf Z}_N)$ and are different from
the ones in our case. A drawback of that setup is that the trivial
gauge field $A=0$ is excluded; hence one cannot consider, e.g., the free
gluon propagator. Our setup, where the space of gauge fields is unrestricted
and the gauge transformations are instead restricted to $\G_0\,$, avoids this.}
Regarding $\A$ as a $\G_0$-bundle over $\A/\G_0\,$, 
a good $\G_0$ gauge fixing is equivalent to a trivialisation
\be 
\A\;\simeq\A/\G_0\,\times\,\G_0
\label{3.1}
\ee  
Since $\pi_n(\A)=0$, existence of a trivialisation implies that
$\pi_n(\A/\G_0)=0$ and $\pi_n(\G_0)=0$ for all $n\!=\!0,1,2,\dots$. 
Nonexistence of good $\G_0$ gauge fixings now follows from the fact that
there are always nonvanishing $\pi_n(\G_0)$'s, which can be seen as follows.
Since the topological structure of SU($N$) is essentially
$S^3\times{}S^5\times\cdots\times{}S^{2N-1}$ modulo a finite set
of equivalence relations, there are smooth maps 
$\Phi:S^{2n-1}\times{}T^4\to\SU(N)$ with nonvanishing
degree for $1\le{}n\le{}N\!-\!2$. Such maps still exist when a condition
$\Phi(\theta,x_0)\!=\!1\,$ $\forall\theta\in{}S^{2n-1}$ is imposed (see
the Appendix). In the latter case, $\Phi$ corresponds to a map 
$S^{2n-1}\to\G_0\,$, $\theta\mapsto\phi_{\theta}$ where 
$\phi_{\theta}(x)\!=\!\Phi(\theta,x)\,$, i.e. $\Phi$ can be regarded as
an element in $\pi_{2n-1}(\G_0)$. This element is clearly nonzero since
$\Phi$ has nonvanishing degree. 
It follows that $\pi_{2n-1}(\G_0)\ne0$
for $1\le{}n\le{}N\!-\!2\,$, which in turn implies that good $\G_0$
gauge fixings do not exist when $N\ge3$. They don't exist in the $N\!=\!2$
case either, since in that case $\pi_0(\G_0)\ne0$ due to the existence of maps
$T^4\to\SU(2)$ which cannot be continuously deformed to the identity map.

These considerations do not have an immediate lattice counterpart
because $\pi_n(\U)\ne0$ for the space $\U$ of lattice gauge fields.
Therefore, we now present another perspective on the above obstructions
to good $\G_0$ gauge fixings by showing that they are the obstructions
to trivialising the $\G_0$-bundle $\A$ over certain $2n$-spheres in
$\A/\G_0$ in the $N\ge3$ case, and the obstruction to trivialising $\A$ over
certain circles in $\A/\G_0$ in the $N\!=\!2$ case.

A smooth map $\Phi:S^{m-1}\times{}T^4\to\SU(N)$ with 
$\Phi(\theta,x_0)\!=\!1\,$ $\forall\theta\in{}S^{m-1}$, together with a 
gauge field $A\in\A$, determines an $m$-sphere in $\A/\G_0$ as follows. 
Define the $S^{m-1}$ family $\phi_{\theta}$ in $\G_0$ by 
$\phi_{\theta}(x)=\Phi(\theta,x)\,$; then an $S^{m-1}$-family in $\A$ is 
obtained by setting $A^{\theta}=\phi_{\theta}\cdot{}A$. Extend this to
a $B^{m}$ family by setting $A^{(\theta,t)}=tA^{\theta}$. Here $B^{m}$
denotes the $m$-dimensional unit ball and $t$ is the radial coordinate.
Since the $A^{(\theta,1)}$'s are all gauge equivalent, the $B^{m}$ family
descends to an $S^{m}$ family in the orbit space, i.e. an $m$-sphere 
in $\A/\G_0$ which we denote by $\S^{m}$.\footnote{We are assuming that
no two interior points in the $B^{m}$ family $A^{(\theta,t)}$ are gauge
equivalent, which will be true in the generic case.}
In the $m\!=\!1$ case, $S^{m-1}\!=\!S^0$ is to be regarded as the boundary
of $B^1$, i.e. the disjoint union of two points. In this case, $\Phi$
consists of two maps $\phi,\phi{}':T^4\to\SU(N)$.

\vspace{1ex}

\noindent {\bf Proposition}. 
The $\G_0$-bundle $\A$ is trivialisable over $\S^{m}$ if and only if
$\Phi$ can be extended to a smooth map $\tPhi:B^{m}\times{}T^4\to\SU(N)$
with $\tPhi(\theta,t,x_0)\!=\!1\,$ $\forall\,(\theta,t)$.

\vspace{1ex}

\noindent {\bf Corollary}. (i) $\A$ is not trivialisable over $\S^{m}$ when
the degree of $\Phi$ is nonzero. 
Thus for $N\ge3$ there are $2n$-spheres in $\A/\G_0$ over which $\A$ is
nontrivialisable. \hfill\break
(ii) In the SU(2) case there are circles in $\A/\G_0$ over which $\A$
is nontrivialisable.

\vspace{1ex} 

\noindent {\bf Proof}. Part (i) of
the corollary follows from the proposition by noting
that an extension $\tPhi$ of $\Phi$ corresponds to a smooth family
$\Phi_t:S^{m-1}\times{}T^4\to\SU(N)$, given by $\Phi_t(\theta,x)=
\tPhi(\theta,t,x)$, which describes a smooth deformation of $\Phi=\Phi_1$
to a map $\Phi_0:S^{m-1}\times{}T^4\to\SU(N)$ which is independent of 
$\theta\in{}S^{m-1}$. The degree of such $\Phi_0$ vanishes; it follows 
that the same must be true for all $\Phi_t$ and in particular for $\Phi$. 
But we have already noted above that $\Phi$'s with nonvanishing degree exist
when $m\!=\!2n$ with $1\le{}n\le{}N\!-\!2$. Part (ii) of the corollary 
follows from the $m\!=\!1$ case of the proposition and the fact that there are
maps $\phi:T^4\to\SU(2)$ which cannot be continuously 
deformed to the identity map:
If $\Phi$ is taken to consist of such a $\phi$ and the identity map then
no extension $\tPhi$ connecting these exists.

To prove the Proposition we first show that trivialisability of $\A$
over $\S^{m}$ implies that an extension $\tPhi$ of $\Phi$ exists.
A trivialisation of $\A$ over $\S^{m}\,$,
\be 
\A\Big|_{\S^{m}}\;\simeq\S^{m}\times\G_0\,,
\label{3.2}
\ee
determines a ``gauge-fixed'' $m$-sphere $\S^{m}_f$ in $\A$, defined as the
image of $\S^{m}\times\{1\}$ under the trivialisation map (\ref{3.2}). 
Let $\tA^{(\theta,t)}$ denote the unique element of $\S_f^{m}$ lying in the 
$\G_0$ gauge orbit through $A^{(\theta,t)}$, and let $\rho_{(\theta,t)}$
denote the unique $\G_0$ gauge transformation relating these by
\be
\tA^{(\theta,t)}=\rho_{(\theta,t)}\cdot{}A^{(\theta,t)}
\label{3.3}
\ee
Since $\tA^{(\theta,1)}\equiv\tA$ is independent of $\theta$, we have
$\tA=\rho_{(\theta,1)}\cdot{}A^{(\theta,1)}=
\rho_{(\theta,1)}\phi_{\theta}\cdot{}A\,$, hence 
$\rho\equiv\rho_{(\theta,1)}\phi_{\theta}$ is independent of $\theta$.
Set $\tPhi(\theta,t,x)=\rho_{(\theta,t)}(x)^{-1}\rho(x)$, then 
$\tPhi(\theta,1,x)=\phi_{\theta}(x)=\Phi(\theta,x)$ so $\tPhi$ is an extension
of $\Phi$ with $\tPhi(\theta,t,x_0)\!=\!1\,$ $\forall\,(\theta,t)$.
Conversely, given an extension $\tPhi$ of $\Phi$ with 
$\tPhi(\theta,t,x_0)\!=\!1\,$ $\forall\,(\theta,t)$, define the smooth 
$B^{m}$ family $\rho_{(\theta,t)}$ in $\G_0$ by 
$\rho_{(\theta,t)}(x)=\tPhi(\theta,t,x)^{-1}$, and define $\tA^{(\theta,t)}$
by (\ref{3.3}). From the definitions, 
$\rho_{(\theta,1)}(x)\phi_{\theta}(x)\!=\!1$ which implies that 
$\tA^{(\theta,1)}$ is independent of $\theta$ and therefore that the family
$\tA^{(\theta,t)}$ is an $m$-sphere $\S_f^{m}$ in $\A$. It is clear from the
constructions that the $\G_0$ orbit through $A^{(\theta,t)}$ intersects 
$\S_f^{m}$ at precisely one point, namely $\tA^{(\theta,t)}$, so $\S_f^{m}$
is a ``gauge-fixed'' submanifold for $\A|_{\S^{m}}$ determining 
a trivialisation (\ref{3.2}). \qed

In preparation for the lattice considerations in the next section
we conclude this section with an alternative ``fermionic'' proof of the 
Corollary of the Proposition above,
based on families index theory for the Dirac operator coupled to gauge 
fields \cite{AS} and Witten's global SU(2) anomaly \cite{Witten}.
A well-known fact, following from the results of \cite{AS}, is that the
topological charge (integrated Chern character) of the index bundle of the
Dirac operator over the above $2n$-sphere $\S^{2n}$ in $\A/\G_0$ 
equals the degree of $\Phi$.
If $\S^{2n}$ is contractible in $\A/\G_0$ then the topological charge must
vanish, so the $\Phi$'s with nonvanishing degree determine noncontractible
$2n$-spheres in $\A/\G_0\,$, in which case $\pi_{2n}(\A/\G_0)\ne0$.
On the other hand, as noted earlier, if a good $\G_0$ gauge fixing exists
then $\pi_m(\A/\G_0)=0$ for all $m$. Thus we see again that the degree of
$\Phi$ is an obstruction to the existence of good $\G_0$ gauge fixings. 

The fermionic proof that the degree of $\Phi$ is an obstruction to 
trivialisability of $\A$ over $\S^{2n}$ (part (i) of the Corollary)
is as follows. If such a trivialisation exists then the corresponding
``gauge fixed'' $2n$-sphere $\S_f^{2n}$ in $\A$ is related to $\S^{2n}$
as in (\ref{3.3}) by a $B^{2n}$ family $\rho_{(\theta,t)}$ of $\G_0$ 
gauge transformations. Recalling that the index bundle over $\A/\G_0$
is obtained from the index bundle over $\A$ by identifying the fiber
over $A$ with the fibre over $\phi\cdot{}A$ via the action of $\phi$
on the space of spinor fields, it is easy to see that the family
$\rho_{(\theta,t)}$ gives an isomorphism between the restriction of the index
bundle over $\A$ to $\S_f^{2n}$ and the restriction of the index bundle
over $\A/\G_0$ to $\S^{2n}$. Hence the topological charges of these 
restricted bundles coincide, so the topological charge of the index bundle
over $\S_f^{2n}$ is $deg(\Phi)$, implying that $\S_f^{2n}$ is noncontractible
in $\A$ if $deg(\Phi)\ne0$. Since all $2n$-spheres in $\A$ are contractible,
this is a fermionic way to see that $\A$ is not trivialisable over 
$\S^{2n}$ when the degree of $\Phi$ is nonzero.

Turning now to the SU(2) case (part (ii) of the Corollary), we give a
fermionic proof that $\A$ is nontrivialisable over the circle $\S^1$
in $\A/\G_0$ coming from the $B^1$-family $A^t=(1\!-\!t)A+t\phi\cdot{}A$
in $\A$ when $\phi:T^4\to\SU(2)$ is a topologically nontrivial element 
in $\G_0$.\footnote{If necessary we can replace 
$\phi\to\phi(x_0)^{-1}\phi$ to get a topologically nontrivial element 
in $\G_0$.}
Let $\{\lambda_j\}$ denote the positive eigenvalues of the Dirac operator
$\sd^A\,$, and $\{\lambda_j(t)\}$ the flows of these eigenvalues when the 
Dirac operator is coupled to $A^t$.\footnote{The argument requires
$A$ to be chosen such that the Dirac operator coupled to $A$ doesn't have any 
accidental zero-modes.} Let $\N$ denote the net number of crossings of the 
origin by these eigenvalues (counted with sign)
as $t$ increases from 0 to 1. A trivialisation
of $\A$ over $\S^1$ determines a ``gauge-fixed'' circle $\S_f^1$ in $\A\,$,
related to $\S^1$ similarly to (\ref{3.3}) by $\tA^t=\rho_t\cdot{}A^t$
for some family $\rho_t$ of $\G_0$ gauge transformations. Then the 
Dirac operators coupled to $A^t$ and $\tA^t$ have the same eigenvalues,
hence the flows $\{\tlambda_j(t)\}$ of the positive eigenvalues
$\{\tlambda_j\!=\!\tlambda_j(0)\}$ of the Dirac operator coupled to 
$\tA\!=\!\tA^0$ coincide with $\{\lambda_j(t)\}\,$, and the net number
$\N_f$ of crossings of the origin by $\{\tlambda_j(t)\}$ coincides with $\N$. 
Clearly $\N_f$ can only change by an even integer under a deformation of the 
circle $\S_f^1$ in $\A$. Therefore, if $\S_f^1$ is contractible then
$\N_f=0$ (mod 2), while on the other hand $\N_f=1$ (mod 2) implies $\S^1$ is 
noncontractible in $\A$. Since all circles in $\A$ are contractible, the
former must hold. It follows that $\A$ is not trivialisable over $\S^1$
when $\phi$ is topologically nontrivial, since in this case it is known
that $\N=1$ (mod 2) ---this is Witten's global SU(2) anomaly 
\cite{Witten}.

\section{Lattice considerations}

In this section we describe a lattice version of the continuum considerations
of Section 4. 
A good $\G_0$ gauge fixing in the SU($N$) lattice gauge theory is equivalent 
to a trivialisation
\be
\U\;\simeq\;\U/\G_0\times\G_0
\label{4.3}
\ee
The $\pi_n(\U)$'s have not been determined at present, and can be 
nonvanishing (as we already know from Section 2). 
Hence the initial considerations of 
Section 4 do not have an immediate lattice counterpart. The subsequent parts of
Section 4 do have lattice counterparts though, as we now discuss.
A map $\Phi:S^{m-1}\times\{\mbox{lattice sites}\}\to\SU(N)$ with 
$\Phi(\theta,x_0)\!=\!1\,$ $\forall\theta\in{}S^{m-1}\,$, together with
a gauge field $U\in\U$, determine an $S^{m-1}$-family $\phi_{\theta}$
in $\G_0$ where $\phi_{\theta}(x)\!=\!\Phi(\theta,x)$, and an $S^{m-1}$-family 
$U^{\theta}=\phi_{\theta}\cdot{}U$ in $\U$. An extension of $U^{\theta}$
to a $B^{2n}$-family $U^{(\theta,t)}$ then gives an $m$-sphere $\S^m$
in $\U/\G_0$. Such an extension need not exist in general; nevertheless
it is expected that generic $m$-spheres in $\U/\G_0$ will arise in this way.
Examples of such $\S^m$ are given 
by the lattice transcripts of the $m$-spheres in $\A/\G_0$ discussed
in Section 4, i.e. $\Phi$, $U$ and $U^{(\theta,t)}$ are the lattice 
transcripts of $\Phi$, $A$ and $A^{(\theta,t)}\,$; the resulting family 
$U^{(\theta,t)}$ is guaranteed to lie in $\U$ when the lattice is sufficiently 
fine, cf. the discussion following (\ref{2}) above.

Now, an argument completely analogous to the proof 
of the Proposition of Section 4
shows that the $\G_0$-bundle $\U$ is trivialisable
over $\S^m$ if and only if $\Phi$ can be smoothly extended to a map
$\tPhi:B^m\times\{\mbox{lattice sites}\}\to\SU(N)$ with 
$\tPhi(\theta,t,x_0)\!=\!1\,$ $\forall\,(\theta,t)$. 
Since trivialisations of $\U$ (i.e. good $\G_0$ gauge fixings) are already 
known to exist, this implies that the extension $\tPhi$ is always guaranteed 
to exist. This is in contrast to the continuum case; the difference is due to
the fact that
on the lattice the is no requirement that $\Phi(\theta,x)$ and 
$\tPhi(\theta,t,x)$ be smooth in the $x$-variable, since the lattice sites
are discrete. (Note that the requirement $\tPhi(\theta,t,x_0)\!=\!1$
can always be satisfied in the lattice case by making this part of the
definition of $\tPhi$.) 

The preceding does not mean that every map 
$\Phi:S^{m-1}\times\{\mbox{lattice sites}\}\to\SU(N)$ can by extended to a map 
$\tPhi:B^m\times\{\mbox{lattice sites}\}\to\SU(N)$. 
Clearly this is not possible, since, in cases where $m$ is even, there are 
maps $S^{m-1}\to\SU(N)\,$, $\theta\to\Phi(\theta,x)$ which wrap $S^{m-1}$
nontrivially around an $(m\!-\!1)$-sphere in SU($N$), and therefore cannot
be extended to maps $B^m\to\SU(N)$. What the preceding implies is that
a {\em necessary condition} for the family $U^{\theta}$ in $\U$, constructed 
using $\Phi$, to admit an extension $U^{(\theta,t)}$ is that $\Phi$ admits an
extension $\tPhi$. I.e. the $\Phi$ cannot produce an $m$-sphere $\S^m$ 
in $\U/\G_0$ in the way described above if it doesn't admit an extension
$\tPhi$. 

When $\S^m$ is the lattice transcript of an $m$-sphere in $\A/\G_0$
of the kind discussed in Section 4, i.e. $U^{(\theta,t)}$ is the lattice
transcript of $A^{(\theta,t)}$, the existence of an extension $\tPhi$ for
the lattice transcript $\Phi:S^{m-1}\times\{\mbox{lattice sites}\}\to\SU(N)$ 
can be seen explicitly as follows. Consider the element
in $\pi_{m-1}(\SU(N))$ represented by the map $\theta\mapsto\Phi(\theta,x)$.
Since the continuum $\Phi(\theta,x)$ depends smoothly on $x$, this element
is independent of $x$ (since $T^4$ is connected). It is zero when 
$x\!=\!x_0$ (since $\Phi(\theta,x_0)\!=\!1\,$ $\forall\theta$), and is 
therefore zero for all $x$.
Hence an extension $B^m\to\SU(N)\,$, $(\theta,t)\to\tPhi(\theta,t,x)$
exists for each lattice site $x$. 

We now give the promised second proof of the Theorem of Section 2.  
Unlike the first proof, it does not rely on the existence of of good
$\G_0$ gauge fixings on the lattice, and provides a more explicit description
of noncontractible $2n$-spheres/circles in $\U$. 
Let $\S^{2n}$ be the lattice
transcript of a $2n$-sphere in $\A/\G_0$ of the kind considered in Section 4.
For each lattice site $x$ choose a smooth extension of the map
$S^{2n-1}\to\SU(N)\,$, $\theta\mapsto\Phi(\theta,x)$ to a map
$B^{2n}\to\SU(N)\,$, $(\theta,t)\mapsto\tPhi(\theta,t,x)$ (as discussed in the 
preceding).
In particular, set $\tPhi(\theta,t,x_0)\!=\!1$. This determines a smooth map
$\tPhi:S^{2n-1}\times\{\mbox{lattice sites}\}\to\SU(N)$ with 
$\tPhi(\theta,1,x)=\Phi(\theta,x)$ and $\tPhi(\theta,t,x_0)=1$.
Now define a $B^{2n}$-family $\rho_{(\theta,t)}$ of $\G_0$ lattice gauge 
transformations by $\rho_{(\theta,t)}(x)=\tPhi(\theta,t,x)^{-1}$ and set
\be
U_f^{(\theta,t)}=\rho_{(\theta,t)}\cdot{}U^{(\theta,t)}
\label{4.4}
\ee
An easy consequence of the definitions is that $U_f^{(\theta,1)}=U$ 
independent of $\theta\in{}S^{2n-1}$. Hence the family $U_f^{(\theta,t)}$
is actually a $2n$-sphere $\S_f^{2n}$. Since $U_f^{(\theta,t)}$ is gauge
equivalent to $U^{(\theta,t)}\,$, this $2n$-sphere is guaranteed to be 
contained in $\U$, and to satisfy the admissibility condition (\ref{2})
for any $\epsilon>0$, when the lattice spacing is sufficiently small.
In the following we apply the families index theory for the overlap Dirac 
operator to show that $\S_f^{2n}$ is noncontractible in $\U$ when $deg(\Phi)$
is nonvanishing and the lattice is sufficiently fine. Since
$\Phi$'s with nonzero degree exist when $1\le{}n\le{}N\!-\!2$, this will 
establish the first part of the Theorem.

A formula for the topological charge $Q_{2n}$ of the index bundle of the 
overlap Dirac operator over the $2n$-sphere $\S^{2n}$ in $\U/\G_0$
was derived in \cite{DA(I)}. It reduces in the classical continuum limit to
$deg(\Phi)$, the topological charge
of the corresponding $2n$-sphere in $\A/\G_0$ \cite{DA(I),DA(II)}.
Thus $Q_{2n}=deg(\Phi)$ when the lattice spacing is sufficiently small.
We now note that, just as in the continuum situation in Section 4,
the relation (\ref{4.4}) between $\S^{2n}$ and the $2n$-sphere $\S_f^{2n}$
in $\U$ implies that the index bundle over $\S^{2n}$ is isomorphic
to the index bundle over $\S_f^{2n}\,$, which in turn implies that their
topological charges are the same. To see this explicitly, recall from
\cite{DA(I)} that, for $n\ge1$, the topological charges of the index bundle
over $\S^{2n}$ and $\S_f^{2n}$ coincide with those of the vectorbundle
$\hC$, given as follows.\footnote{This bundle was denoted ``$\hC_+$'' in
\cite{DA(I)} but we omit the ``$+$'' subscript here.} 
The fibre of $\hC$ over $U\in\U$ is
$\hC^U=P^U(\C)$ where $P^U$ is a projection operator 
acting on the space $\C$ of lattice spinor fields, given by
\be
P^U={\textstyle \frac{1}{2}}(1+\hg5^U)\quad,\qquad\,\hg5^U=\g5(1-aD^U)
\label{4.4a}
\ee
where $D^U$ is the overlap Dirac operator coupled to $U$.
Set $\hC_f^{(\theta,t)}:=\hC^{U_f^{(\theta,t)}}\,$, then 
\be
\hC\Big|_{\S_f^{2n}}=\{\;\hC_f^{(\theta,t)}\;\}
\label{4.5}
\ee
Gauge covariance of $P^U$ implies $\phi(\hC^U)=\hC^{\phi\cdot{}U}$ for all
$\phi\in\G_0\,$, $\U\in\U\,$, so $\hC$ descends to a vector bundle over
$\U/\G_0$. The restriction of this bundle to $\S^{2n}$ is
\be
\hC\Big|_{\S^{2n}}=\{\;\hC^{(\theta,t)}\;\}/_{\sim}
\label{4.6}
\ee
where $\hC^{(\theta,t)}:=\hC^{U^{(\theta,t)}}$ and the equivalence relation
$\sim$ means identify each $\hC^{(\theta,1)}$ with $\hC^{(0,1)}$ via the
isomorphism
\be
\phi_{\theta}:\hC^{(0,1)}\stackrel{\simeq}{\to}\hC^{(\theta,1)}
\label{4.7}
\ee
For $t<1$ the fibres $\hC^{(\theta,t)}$ and $\hC_f^{(\theta,t)}$ are 
isomorphic via the gauge transformation $\rho_{(\theta,t)}$ in (\ref{4.4}).
This extends to a well-defined isomorphism between the fibres at 
$t\!=\!1$ since, as an easy consequence of the definitions, 
$\rho_{(\theta,1)}\phi_{\theta}=1$ independent of $\theta$, i.e.
$\rho_{(\theta,1)}$ respects the equivalence relation in (\ref{4.6}).
Thus $\hC|_{\S^{2n}}$ and $\hC|_{\S_f^{2n}}$ are isomorphic, and therefore have
the same topological charge $Q_{2n}$ as claimed.
It follows that if $\S_f^{2n}$ is contractible in $\U$ then $Q_{2n}=0$.
Therefore, if $deg(\Phi)\ne0$ then the $2n$-sphere $\S_f^{2n}$ constructed
above must be noncontractible when the lattice is sufficiently fine.
This proves the first part of the theorem.

The proof of the remaining part of the theorem is as follows.
Let $\S^1$ be the lattice transcript of a circle
in $\A/\G_0$ of the kind discussed in Section 3, i.e. start with a 
$\phi\in\G_0$, pick $A\in\A\,$, then the lattice transcript $U^t$ of the
family $A^t=(1\!-\!t)A\!+\!t\phi\cdot{}A$ determines a circle $\S^1$ in
$\U/\G_0$. For each lattice site $x$ choose a smooth path $\phi_t(x)$ in 
SU(2) connecting the identity element $\phi_0(x)=1$ to $\phi(x)$. 
In particular, set $\phi_t(x_0)=1\,$ $\forall\,t$. Define the family
$\rho_t$ in $\G_0$ by $\rho_t(x)=\phi_t(x)^{-1}$ and set
\be 
U_f^t=\rho_t\cdot{}U^t\,.
\label{4.8}
\ee
Since $\rho_1\phi=1=\rho_0\,$, implying $U_f^0=U_f^1\,$, the family $U_f^t$
is a circle $\S_f^1$. Just as in the case of $\S_f^{2n}$ it is 
guaranteed to lie in $\U$ when the lattice is sufficiently fine.
The fermionic argument of Section 3 in the SU(2) case now has the following 
lattice version.
The overlap Dirac operator $D$ is normal and hence has a complete set of
eigenvectors. The eigenvalues lie on a circle in the complex plane which 
passes through the origin, and the nonreal eigenvalues come in complex
conjugate pairs.\footnote{The convention used in defining the overlap Dirac 
operator is that the $\gamma^{\mu}$ matrices are hermitian. This corresponds
in the continuum to antihermitian Dirac operator with purely imaginary 
eigenvalues, with the nonzero ones coming in complex conjugate pairs.}
Let $\{\lambda_j\}$ be the eigenvalues of $D$ coupled to $U^0$ which have
${\rm Im}(\lambda)>0\,$, and $\{\lambda_j(t)\}$ the flows of these eigenvalues
when $D$ is coupled to $U^t$. Let $\N$ denote the net number of crossings of
the origin (counted with sign) as $t$ increases from 0 to 1. Let $\N_f$
denote the analogous number for $D$ coupled to $U_f^t$. Then, just as in the 
continuum setting, the gauge covariance of $D$ and the relation (\ref{4.8})
imply $\N_f=\N$. Clearly $\N_f$ can only change by an even integer under
a deformation of the circle $\S_f^1$ in $\U$. Therefore, if $\S_f^1$ is 
contractible in $\U$ then $\N_f$ (mod 2) $=0$. On the other hand,
the results of \cite{Bar} show that $\N$ (mod 2) reproduces Witten's global 
SU(2) anomaly in the classical continuum limit.\footnote{The presence 
of the Witten anomaly on the lattice has also been verified numerically in 
\cite{Neu(SU(2)),Bar1}.} It follows that $\S_f^1$
is noncontractible $\U$ when $\phi$ is topologically nontrivial
and the lattice is sufficiently fine. 
This completes our second proof of the Theorem of Section 2.

\vspace{1ex}

\noindent {\bf Remarks}. (i) The noncontractible $2n$-spheres $\S_f^{2n}$
and circles $\S_f^1$
in $\U$ constructed in the preceding are ``rough'': they cannot arise as
lattice transcripts of smooth continuum $2n$-spheres and circles in $\A$
since the contractibility of the latter implies contractibility of 
their lattice transcripts when the lattice is sufficiently fine.
The roughness of $\S_f^{2n}$ and $\S_f^1$ 
originates from the roughness of the of the extensions 
$\tPhi:B^{2n}\times\{\mbox{lattice sites}\}\to\SU(N)$
and $\widetilde{\phi}:B^1\times\{\mbox{lattice sites}\}\to\SU(2)$ 
(where $\widetilde{\phi}(t,x)\!=\!\phi_t(x)\,$);
these do not have smooth continuum versions when $deg(\Phi)\ne0$, and 
$\phi$ is topologically nontrivial, respectively. \hfill\break
(ii) Since the continuum gauge field $A$ used as part of the starting point
for constructing $S_f^{2n}$ and $\S_f^1$ in the preceding is smooth and 
continuous on $T^4$ and therefore has vanishing topological charge, 
$\S_f^{2n}$ and $\S_f^{2n}$ lie in the trivial topological sector of $\U$.
However, noncontractible $2n$-spheres and circles in the other topological
sectors are readily constructed along the same lines as above by starting
with a topologically nontrivial $A$ in a singular gauge, such that $A$ is
still continuous on $T^4$ and smooth away from the singularity, and the 
lattices are restricted to those for which the singularity of $A$ doesn't 
lie on a lattice link.

Finally, we point out that the topological charge $Q_{2n}$ of the overlap 
Dirac index bundle over the $2n$-spheres $\S_f^{2n}$ is associated with
a monopole interpretation for a certain canonical connection on $\U$ with
values in the bundle $\hC$. 
The bundle $\hC$ arises as $\hC=\P(\C)$ where $\C$ is to be 
interpreted as the trivial bundle over $\U$ with fibre $\C$ (the space of 
lattice spinor fields on $T^4$) and $\P:\C\to\C$ is an orthogonal projection 
map whose action on the fibres is given by $P^U$, defined in (\ref{4.4a}). 
Then $\hC$ has the canonical connection $\nabla=\P{}d$, where $d$ is the 
exterior derivative on $U$.\footnote{This can be regarded as a generalised 
Levi-Civita connection, cf. appendix B of \cite{Karoubi}: The Levi-Civita
connection of a riemannian manifold $M$ embedded in Euclidean space 
${\bf R}^n$ can be written as $\nabla=\P{}d$ where $P_x$ is the orthogonal
projection of ${\bf R}^n$ onto the tangent space of $M$ at $x$.}
On $\U_{\rm initial}$ we see from (\ref{4.4a}) that $\nabla$ is singular at 
the points $U$ for which the overlap Dirac operator $D^U$, and hence the
fermionic topological charge, are ill-defined. 
Such singularities are present in the interior of any $(2n\!+\!1)$-ball in 
$\U_{\rm initial}$ with $\S_f^{2n}$ as its boundary when $Q_{2n}$ is
nonvanishing. (Such balls always exist since $\S_f^{2n}$ is contractible in 
$\U_{\rm initial}$.)
These singularities of $\nabla$ are monopole-like: 
the topological charge of $\nabla$ on $\S_f^{2n}$, given by integrating the
Chern character of $\nabla$ over $\S_f^{2n}$, equals $Q_{2n}$.
Indeed, the Chern charater of $\nabla$ is a representative for 
the Chern character of the bundle $\hC$, and it was shown in \cite{DA(I)} 
that the nonzero degree parts of the latter coincide with those of
the overlap Dirac index bundle on $\U$.  
(In fact the connection $\nabla$ was used to derive the
formula (``lattice families index theorem'') for the Chern character of
the lattice index bundle in \cite{DA(I)}.)

\section{Summary}

When the decomposition of the space of lattice gauge fields into topological
sectors is specified by the fermionic topological charge, it is to be expected
that fermionic techniques will be required to determine the (topological)
structure of the sectors. In this paper we have seen 
that the lattice families index theory developed in \cite{DA(I),DA(II)}
is a useful tool in this regard. Using it, we obtained first results on
the topology of the sectors of SU($N$) lattice gauge
fields on $T^4\,$: For sufficiently fine lattices, the sectors were shown to
contain noncontractible $2n$-spheres when $1\le{}n\le{}N\!-\!2\,$,
and noncontractible circles in the SU(2) case. These are new topological 
features that arise as a consequence of excluding the lattice gauge fields
for which the fermionic topological charge is ill-defined.

Two proofs of this result were given. The first of these, in Section 2,
was a short, rather implicit argument using results on the topological
structure of the orbit space $\U/\G_0$ obtained previously from the families
index theory for the overlap Dirac operator in \cite{DA(I)},
and exploiting the fact that $\G_0$ gauge fixings without the Gribov problem
exist on the lattice. The second argument, in Section 5, used the lattice
families index theory in a more direct way, without relying on the 
existence of good $\G_0$ gauge fixings. It led to noncontractible
$2n$-spheres $\S_f^{2n}$ in the $N\ge3$ case, and noncontractible circles 
$\S_f^1$, in the $N\!=\!2$ case, obtained by a quite explicit 
prescription as follows: Start with a continuum gauge field $A$
and smooth map $\Phi:S^{2n-1}\times{}T^4\to\SU(N)$ with 
$\Phi(\theta,x_0)\!=\!1\,$ $\forall\,\theta$ and $deg(\Phi)\ne0$. 
Then, for each lattice site $x$, extend the map $S^{2n-1}\to\SU(N)\,$, 
$\theta\mapsto\Phi(\theta,x)$ to a map $B^{2n}\to\SU(N)\,$, 
$(\theta,t)\mapsto\tPhi(\theta,t,x)$. This determines $\S_f^{2n}$ via 
(\ref{4.4}). $\S_f^1$ is similarly determined via (\ref{4.8}) after
starting with a topologically nontrivial map $\phi:T^4\to\SU(2)$ and choosing
for each lattice site $x$ a path $\phi_t(x)$ from $1$ to $\phi(x)$ in SU(2).
The noncontractible $\S_f^{2n}$'s and $\S_f^1$'s are rough -- they do not 
have smooth continuum versions. 
Nevertheless, the gauge fields contained in them satisfy the admissibility
condition (\ref{2}) for any given $\epsilon$ when the lattice is 
sufficiently fine.

The arguments furthermore show that the number
of elements in $\pi_{2n}(\U)$ for $1\le{}n\le{}N\!-\!2$ becomes infinite
in the continuum limit, and that the topological charges of the index bundle
over the $\S_f^{2n}$'s are associated with a monopole interpretation for a
certain canonical connection on the space of lattice gauge fields.

It would be interesting to to see if further results 
on the topology of $\U$ can be extracted via the lattice families index theory
or other fermionic techniques. A number of basic questions remain to be 
answered, for example: Is a given topological sector of $\U$ path-connected,
or can there be more than one connected component? 
(The decomposition (\ref{7}) shows that the answer is affirmative in
the U(1) case, at least when the admissibility condition is imposed.)
Determining the topology of $\U$ would be necessary for extending L\"uscher's
existence proof for gauge invariant abelian chiral gauge theory on the lattice
\cite{L(abelian)} to the SU($N$) case.

In the continuum setting, the topological charge (integrated Chern character)
of the Dirac index bundle over $2n$-spheres in the orbit space $\A/\G_0$ 
was seen to be an obstruction to the existence of good $\G_0$ gauge fixings.
Explicit examples were described where the obstruction is given by the degree
of maps $\Phi:S^{2n-1}\times{}T^4\to\SU(N)\,$ ($1\le{}n\le{}N\!-\!2$) in the
$N\ge3$ case. The Witten global anomaly was seen to be an obstruction to
the existence of good $\G_0$ gauge fixings in the SU(2) case.
In the lattice setting the situation is quite different: Good $\G_0$ gauge
fixings exist, and the topological charge of the overlap Dirac index bundle
over the $2n$-spheres $\S^{2n}$ in $\U/\G_0$ (the lattice transcripts of the 
above-mentioned $2n$-spheres in $\A/\G_0$) 
was seen to be an obstruction to the 
contractibility of the $2n$-spheres $\S_f^{2n}$ in $\U$ obtained from
$\S^{2n}$ via a trivialisation of the $\G_0$-bundle $\U$ over $\S^{2n}$. 
Similarly, the lattice version of the Witten global anomaly was seen to be an 
obstruction to contractibility of the circles $\S_f^1$ in $\U$ obtained from
$\S^1$ in the SU(2) case. Thus the obstructions to the contractibility of
the $2n$-spheres/circles in $\U$ are the lattice counterparts of the 
obstructions to good $\G_0$ gauge fixings in the continuum setting. 
The correspondence is made even stronger by the fact that the lattice
obstructions coincide with the continuum ones for sufficiently fine lattices
by the classical continuum limit results of \cite{DA(I)} and \cite{Bar}.

\vspace{1ex}

{\bf Acknowledgements}. I thank Ting-Wai Chiu and his students for many
stimulating discussions and kind hospitality at NTU. I would also like to
thank Kazuo Fujikawa, Masato Ishibashi and Yoshio Kikukawa for discussions 
and kind hospitality during a visit to Japan, and Herbert Neuberger for his
encouraging interest in the preliminary version of this work at the Cairns
workshop. At NTU the author was supported by the Taiwan NSC (grant numbers
NSC89-2112-M-002-079 and NSC90-2811-M-002-001).

\section*{Appendix}

\noindent{\bf A remark on the condition $\Phi(\theta,x_0)\!=\!1\,$ 
$\forall\,\theta$}

\vspace{1ex}

For $1\le{}n\le{}N\!-\!2$ there are maps $\Phi:S^{2n-1}\times{}T^4\to\SU(N)$
with arbitrary nonvanishing degree. In the following we point out that
such maps continue to exist when the condition $\Phi(\theta,x_0)=1\,$
$\forall\,\theta\in{}S^{2n-1}$ is imposed, with $x_0$ being some arbitrary
basepoint in $T^4$. Start with a map $\Phi:S^{2n+3}\to\SU(N)$ with arbitrary
degree $d$, obtained by wrapping the $S^{2n+3}\,$ $d$ times around the
$2n\!+\!3$-sphere in $\SU(N)$. Choose a point $p_0$ in $S^{2n+3}$ and 
redefine $\Phi(p)\to\Phi(p_0)^{-1}\Phi(p)$ so that $\Phi(p_0)=1$.
Now view $S^{2n+3}$ as the box $\lb0,1\rb^{2n+3}$ with all boundary points 
identified with $p_0$. Then $\Phi$ can be viewed as a map 
$\Phi:\lb0,1\rb^{2n+3}\to\SU(N)$ which maps all boundary points of the box
to $1$. View $\lb0,1\rb^{2n+3}$ as $\lb0,1\rb^{2n-1}\times\lb0,1\rb^4$.
If $x_0$ is a boundary point in $\lb0,1\rb^4$ then $(u,x_0)$ is a boundary 
point in $\lb0,1\rb^{2n+3}$ for all $u\in\lb0,1\rb^{2n-1}\,$, hence
$\Phi(u,x_0)=1\,$ $\forall\,u$. Now impose periodic boundary conditions
on $\lb0,1\rb^4$ to get $T^4$, and identify all boundary points in 
$\lb0,1\rb^{2n-1}$ to get $S^{2n-1}$. The map 
$\Phi:\lb0,1\rb^{2n-1}\times\lb0,1\rb^4\to\SU(N)$ respects these 
identifications (since it maps all boundary points to $1$) and can therefore
be regarded as a map $\Phi:S^{2n-1}\times{}T^4\to\SU(N)$. This map satisfies
$\Phi(\theta,x_0)=1\,$ $\forall\,\theta$ and has the same degree $d$ as the
original $\Phi$. 

In light of this, the obstructions to gauge invariance of the chiral fermion
determinant mentioned in Section 2, which are given by the degree of some
$\Phi$, continue to be present when $\G$ is restricted to $\G_0$.


\begin{thebibliography}{XXX}



\bibitem{ov}
R. Narayanan and H. Neuberger, Phys. Lett. B {\bf 302}, (1993) 62;
%%CITATION = HEP-LAT 9212019;%%
Phys. Rev. Lett {\bf 71}, (1993) 3251; 
%%CITATION = HEP-LAT 9308011;%%
Nucl. Phys. B {\bf 412}, (1994) 574; 
%%CITATION = HEP-LAT 9307006;%%
Nucl. Phys. B {\bf 443}, (1995) 305 
%%CITATION = HEP-TH 9411108;%%

\bibitem{Neu1}
H. Neuberger, Phys. Lett. B {\bf 417}, (1998) 141  
%%CITATION = HEP-LAT 9707022;%%

\bibitem{L(abelian)}
M. L\"uscher, Nucl. Phys. B {\bf 549}, (1999) 295 
%%CITATION = NUPHA,B549,295;%%

\bibitem{Neu(noncompactU(1))}
H. Neuberger, Phys. Rev. D {\bf 63}: 014503 (2001)
%%CITATION = HEP-LAT 0002032;%%

\bibitem{Suzuki(general)}
T. Fujiwara, H. Suzuki and K. Wu, Phys. Lett. B {\bf 463} (1999) 63;
%%CITATION = HEP-LAT 9906016;%%
Nucl. Phys. B {\bf 569} (2000) 643
%%CITATION = HEP-LAT 9906015;%%

\bibitem{Kiku}
Y. Kikukawa and Y. Nakayama, Nucl. Phys. B {\bf 597} (2001) 519
%%CITATION = NUPHA,B597,519;%%

\bibitem{Suzuki}
H. Suzuki, Nucl. Phys. B {\bf 585} (2000) 471
%%CITATION = NUPHA,B585,471;%%

\bibitem{L(pert)}
M. L\"uscher, JHEP {\bf 0006}:028 (2000)
%%CITATION = HEP-LAT 0006014;%%

\bibitem{L(nonabelian)}
M. L\"uscher, Nucl. Phys. B {\bf 568}, (2000) 162 
%%CITATION = NUPHA,B568,162;%%

\bibitem{Suzuki(real)}
H. Suzuki, JHEP {\bf 0010}:039 (2000)
%%CITATION = HEP-LAT 0009036;%%

\bibitem{DA(I)}
D.H. Adams, Nucl. Phys. B {\bf 624} (2002) 469
%%CITATION = NUPHA,B624,469;%%

\bibitem{DA(II)} 
D.H. Adams, {\em Families index theory for Overlap lattice Dirac 
operator. II} (in preparation)

\bibitem{Bar}
O. B\"ar and I. Campos, Nucl. Phys. B {\bf 581}, (2000) 499 
%%CITATION = NUPHA,B581,499;%%

\bibitem{Creutz}
M. Creutz, Phys. Rev. D {\bf 15}, (1977) 1128
%%CITATION = PHRVA,D15,1128;%%

\bibitem{DeTar}
C. DeTar, J.E. King, S.P. Li and L. McLerran, Nucl. Phys. B {\bf 249} 
(1985) 621
%%CITATION = NUPHA,B249,621;%%

\bibitem{DA(JMP)}
D.H. Adams, J. Math. Phys. {\bf 42} (2001) 5522
%%CITATION = HEP-LAT 0009026;%%

\bibitem{L(local)}
P. Hern\'andez, K. Jansen and M. L\"uscher, Nucl. Phys. B 
{\bf 552}, (1999) 363 
%%CITATION = NUPHA,B552,363;%%

\bibitem{Neu(bound)}
H. Neuberger, Phys. Rev. D {\bf 61}, 085015 (2000)
%%CITATION = HEP-LAT 9911004;%%

\bibitem{Gribov}
V.N. Gribov, Nucl. Phys. B {\bf 139} (1978) 1
%%CITATION = NUPHA,B139,1;%%

\bibitem{Neu(SU(2))}
H. Neuberger, Phys. Lett. B {\bf 437}, (1998) 117 
%%CITATION = PHLTA,B437,117;%%

\bibitem{Bar1}
O. B\"ar and I. Campos, Nucl. Phys. (Proc. Suppl.) {\bf 83}, (2000) 594 
%%CITATION = HEP-LAT 9909081;%%

\bibitem{Mandula}
J.E. Mandula and M.C. Ogilvie, Phys. Rev. D {\bf 41}, (1990) 2586
%%CITATION = PHRVA,D41,2586;%%

\bibitem{lor}
F. Lenz, J.W. Negele, L. O'Raifeartaigh and M. Thies, Ann Phys. {\bf 258}
(2000) 25
%%CITATION = HEP-TH 0004200;%%

\bibitem{gfreview}
L. Giusti, M.L. Paciello, C. Parrinello and S. Petrarca,
Int. J. Mod. Phys. A {\bf 16} (2001) 3487
%%CITATION = HEP-LAT 0104012;%%

\bibitem{vBaal}
P. van Baal, in Nato Advanced Study Institute on
Confinement, duality, and nonperturbative aspects of QCD
(Cambridge, 1997), ed: P. van Baal (Plenum Press, New York, 1998), 
hep-th/9711070
%%CITATION = HEP-TH 9711070;%%

\bibitem{Narayanan-Kiku(overlap)}
R. Narayanan, Phys. Rev. D {\bf 58} (1998) 097501;
%%CITATION = HEP-LAT 9802018;%%
Y. Kikukawa and A. Yamada, hep-lat/9810024
%%CITATION = HEP-LAT 9810024;%%

\bibitem{AG}
L. Alvarez-Gaum\'e and P. Ginsparg, Nucl. Phys. B {\bf 243}, (1984) 449   
%%CITATION = NUPHA,B243,449%%

\bibitem{Neu(geom)}
H. Neuberger, Phys. Rev. D {\bf 59}, 085006 (1999)
%%CITATION = HEP-LAT 9802033;%%

\bibitem{DA(NPB)}
D.H. Adams, Nucl. Phys. B {\bf 589}, (2000) 633 
%%CITATION = NUPHA,B589,633;%%

\bibitem{DA(PRL)}
D.H. Adams, Phys. Rev. Lett {\bf 86}, (2001) 200 
%%CITATION = HEP-LAT 9910036;%%

\bibitem{Singer}
I.M. Singer, Comm. Math. Phys. {\bf 60} (1978) 7
%%CITATION = CMPHA,60,7;%%

\bibitem{Syracuse}
G. Jungman, Mod. Phys. Lett. {\bf A7} (1992) 849
%%CITATION = MPLAE,A7,849;%%

\bibitem{AS}
M.F. Atiyah and I.M. Singer, Proc. Natl. Acad. Sci. U.S.A. {\bf 81},
(1984) 2597 
%%CITATION = PNASA,81,2597;%%

\bibitem{Witten}
E. Witten, Phys. Lett. B {\bf 117}, (1982) 324 
%%CITATION = PHLTA,B117,324;%%

\bibitem{Karoubi}
M. Karoubi, Algebraic topology via differential geometry, LMS Lecture Note
Series {\bf 99} (Cambridge Univ. Press, 1987)


\end{thebibliography}
\end{document}